\documentclass[preprint,titlepage,nobibnotes,nofootinbib]{revtex4-2} 
\usepackage{graphicx} 

\usepackage{amsthm}
\usepackage{amsmath,amssymb,amsfonts}
\usepackage{bm}
\usepackage{color}
\usepackage{braket}
\usepackage{listings}
\usepackage{enumerate}
\usepackage{hyperref}
\usepackage{cleveref}
\hypersetup{  
     unicode=false,          
     pdftoolbar=true,        
     pdfmenubar=true,        
     pdffitwindow=true,      
     pdftitle={Prefix Sums via Kronecker Products},
     pdfauthor={Aleksandros Sobczyk, Anastasios Zouzias}, 
     pdfsubject={Parallel Algorithms},   
     pdfcreator={Aleksandros Sobczyk, Anastasios Zouzias},   
     pdfproducer={Aleksandros Sobczyk, Anastasios Zouzias}, 
     pdfkeywords={prefix sum, scan, matrix multiplication, kronecker product}, 
     pdfnewwindow=true,     
     colorlinks=true,       
     linkcolor=blue,        
     citecolor=blue,        
     filecolor=magenta,     
     urlcolor=blue,         
}

\DeclareMathOperator{\vect}{vec}
\DeclareMathOperator{\matt}{mat}

\newcommand{\veca}{\bm{a}}
\newcommand{\vecb}{\bm{b}}
\newcommand{\vecx}{\bm{x}}
\newcommand{\vecy}{\bm{y}}
\newcommand{\vece}{\bm{e}}
\newcommand{\vecz}{\bm{z}}
\newcommand{\vecw}{\bm{w}}

\newcommand{\vecp}{\bm{p}}
\newcommand{\vecg}{\bm{g}}

\newcommand{\vecones}{\bm{1}}

\newcommand{\matA}{\bm{A}}
\newcommand{\matB}{\bm{B}}

\newcommand{\matX}{\bm{X}}
\newcommand{\matY}{\bm{Y}}
\newcommand{\matL}{\bm{L}}
\newcommand{\matU}{\bm{U}}
\newcommand{\matI}{\bm{I}}

\newcommand{\RR}{\mathbb{R}}

\newtheorem{theorem}{Theorem}

\begin{document}
%
\title{Prefix Sums via Kronecker Products\\}
\author{Aleksandros Sobczyk}
\email{aleksandros.sobczyk@h-partners.com}
\author{Anastasios Zouzias}
\email{anastasios.zouzias@huawei.com}
\affiliation{Huawei Technologies Switzerland AG\\}
%
\begin{abstract}
In this work, we revisit prefix sums through the lens of linear algebra.
We describe an identity that decomposes triangular all-ones matrices as a sum of two Kronecker products, and apply it to design recursive prefix sum algorithms and circuits. Notably, the proposed family of circuits is the first one that achieves the following three properties simultaneously: (i) zero-deficiency, (ii) \emph{constant} fan-out per-level, and (iii) depth that is asymptotically strictly smaller than $2\log(n)$ for input length $n$. As an application, we show how to use these circuits to design quantum adders with $1.893\log(n)+O(1)$ Toffoli depth, $O(n)$ Toffoli gates, and $O(n)$ additional qubits, improving the Toffoli depth and/or Toffoli size of existing constructions.
\end{abstract}
%
\maketitle
%
\section{Introduction}
%
Given a sequence of $n$ elements $\vecx(0),\vecx(1),\ldots,\vecx(n-1)$ and an associative binary operator $\circ$, the ``prefix problem'', also known as ``prefix sum'' or ``scan'', is defined as the sequence:
%
\begin{align}
    \vecy(i) = \vecx(0)\circ \vecx(1)\circ\ldots \circ \vecx(i),\quad 0\leq i\leq n-1.
    \label{eq:prefix_sum}
\end{align}
%
Prefix sum is a fundamental concept in algorithm analysis with numerous applications. It plays a central role in the design of arithmetic processors, such as binary adders \cite{adders:sklansky60,adders:brent_kung82} and multipliers \cite{book:scan94}, even on quantum computers \cite{draper2006logarithmic,wang2025optimal}. It is widely used as a building block for parallel graph algorithms  \cite{blelloch_prefix_1990}, matrix arithmetics \cite{spmv:segmv_blelloch93}, and, more recently, language models \cite{gu2024mamba}. 
Due to the wide applicability, prefix sum algorithms have been studied for decades, and their theoretical properties are well-understood. 

The prevailing model to analyze prefix sums is the circuit model; see e.g. \cite{book:scan94}. For a sequence of length $n$, a prefix circuit is a directed acyclic graph where the nodes indicate binary operations $\circ$. The \emph{size}  $S(n)$ is the total number of nodes, the \emph{depth} $D(n)$ is the length of the longest path (number of edges) from an input to an output, and the \emph{fan-out} is the maximum number of outgoing edges of any node. Simplicity of the circuit constructions is also desirable for efficient implementations. It is typically quantified mathematically with the so-called \emph{uniformity} of the circuit family, which measures how much time or space is required by an algorithm to either construct the circuit or to verify its properties.

The simplest way to solve the prefix problem is with a \emph{serial} circuit, which updates each output as $\vecy(i)\gets \vecx(i)\circ \vecy(i-1) $, with size $n-1$ and depth $n-1$. A plethora of sophisticated prefix circuits have been proposed in the literature with logarithmic depth (see also Table~\ref{table:previous_work}). 
Sklansky \cite{adders:sklansky60} described a circuit that achieves the optimal depth $D(n)=\log(n)$, but the size increases to $n\log(n)/2$. Ladner and Fischer \cite{ladner1980parallel} proposed a recursive construction that can reduce the size by slightly increasing the depth. Fich \cite{scan:fich_stoc83} proposed improvements and also proved lower bounds for the size of circuits of minimal depth. Brent and Kung \cite{adders:brent_kung82} described a recursive circuit in the context of parallel adders. 
Snir~\cite{snir1986depth} proved a tight optimality condition for the sum of depth $D(n)$ and size $S(n)$ of any prefix circuit:
%
\begin{align}
    D(n)+S(n) \geq 2n-2.
    \label{eq:snir_bound}
\end{align}
%
Circuit families that satisfy this bound with equality are said to have \emph{zero-deficiency}, and they have attracted significant attention since Snir's discovery. The so-called ``LYD'' family of circuits \cite{lakshmivarahan1987new} improved the depth of Snir's construction while maintaining zero-deficiency, and Lin and Shih further improved the minimum achievable depth \cite{lin1999new}. Lin and co-authors published a series of further developments, focusing on zero-deficiency parallel prefix circuits with bounded fan-out (at most $4$) \cite{lin2003constructing,lin2003z4,lin2004new,lin2024size,lin2005faster}. 
In a landmark work, Zhu, Cheng, and Graham \cite{scan:zero_def_circuits06} provided a simplified proof of Snir's bound, and they also proved that there exist no zero-deficiency circuits with depth less than:
%
\begin{align}
    D_{\min}(n):=\min\{t:F(t)\geq n+1\}-3,
    \label{eq:dmin}
\end{align}
%
where $F(t)$ is the $t$-th Fibonacci number. In addition, they described a circuit family that achieves this lower bound for every $n$. Sheeran and Parberry \cite{sheeran2006new} proposed a circuit family with parametrized fan-out, which yields a circuit with depth $2\log(n)-1$, zero-deficiency, and fan-out two. Lin and Hung \cite{lin2009straightforward} proposed an alternative construction with the same properties. Sergeev \cite{sergeev2024complexity} showed that the $2\log(n)+O(1)$ depth of \cite{sheeran2006new,lin2009straightforward} is essentially optimal for zero-deficiency circuits with fan-out two, and also proposed new constructions that achieve it. 
While the focus of this work is on circuits with bounded fan-in, here we also mention several works that have focused in unbounded fan-in circuits. A seminal result in this literature is that if the underlying semi-group is group free then there there exist constant depth and almost linear size circuits ~\cite{chandra1983unbounded,book:scan94,yeh2000optimal}. For example, in~\cite{book:scan94} it is shown that, for the input length $n$, there exist prefix circuits of depth $4$ and size $2n\log^* (n)$. 
%

\begin{figure}[ht]
  \centering
  \vspace{-8mm}
  \includegraphics[width=0.75\linewidth]{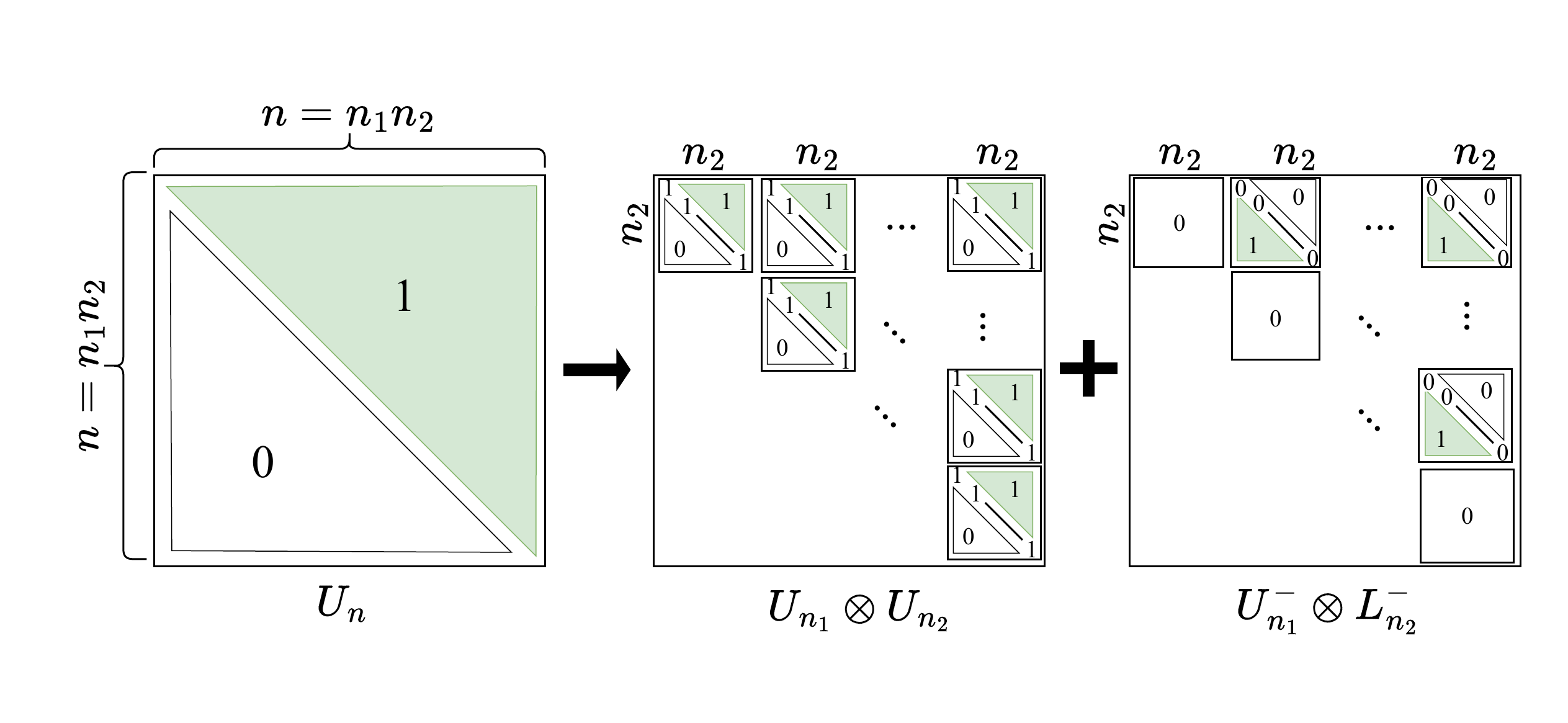}
  \vspace{-8mm}
  \caption{\small Illustration of Kronecker decomposition (Theorem~\ref{theorem:triu_kron}).}
  \label{fig:main_theorem}
\end{figure}

%
\begin{table*}[htb]
    \centering
\caption{Prefix circuits. }
\footnotesize
\setlength{\tabcolsep}{10pt} 
\renewcommand{\arraystretch}{0.9}
\resizebox{\columnwidth}{!}{
    \begin{tabular}{ 
    r r r r l}
    \hline
        Size $S(n)$
        &
        Depth $D(n)$
        &
        Fan-out
        &
        Deficiency
        &
        Reference
    \\\hline\hline
    $n-1$
        &
        $n-1$
        &
        $2$
        &
        $0$
        &
        Serial circuit.$^{(2)}$
    \\ 
        $\tfrac{n}{2}\log(n)$
        &
        $\log(n)$
        &
        $n/2$
        &
        $O(n\log(n))$
        &
        Sklansky \cite{adders:sklansky60}.$^{(1)}$$^{(2)}$
    \\ 
        $n\lceil\log(n)\rceil -n + 1$
        &
        $\log(n)+O(1)$
        &
        $2$
        &
        $O(n\log(n))$
        &
        Kogge--Stone \cite{adders:kogge_stone73},\cite[Sec. 3.1, Fig. 1]{book:scan94}.$^{(2)}$
    \\
        $2n-2$
        &
        $2\log(n)+O(1)$
        &
        $2$
        &
        $O(\log(n))$
        &
        Cyclic Reduction \cite[Sec. 3.1, Fig. 5]{book:scan94}.$^{(2)}$
    \\  
        $2(1+\tfrac{1}{2^k})n-o(n)-k$
        &
        $\log(n)+k$
        &
        $\lfloor\tfrac{n+2^k-1}{2^{k+1}}\rfloor+k$
        &
        $O(n)$
        &
        Ladner--Fischer \cite{ladner1980parallel}, $0\leq k \leq \log(n)$.$^{(1)}$
    \\  
        $2n-\log(n)-2$
        &
        $2\log(n)-1$
        &
        $2$
        &
        $O(\log(n))$
        &
        Brent--Kung \cite{adders:brent_kung82}.$^{(1)}$$^{(2)}$
    \\  
        $2n-2-D(n)$
        &
        $2\log(n)+O(1)$
        &
        $\lceil\log(n)\rceil+1$
        &
        $0$
        &
        Snir \cite{snir1986depth}.
    \\
        $2n-2-D(n)$
        &
        $2\log(n)+O(1)$
        &
        $2\lceil\log(n)\rceil-2$
        &
        $0$
        &
        Lakshmivarahan--Yang--Dhall \cite{lakshmivarahan1987new}, $n\geq 9$.
    \\
        $2n-2-D(n)$
        &
        $2\log(n)+O(1)$
        &
        $\lceil \log(n)\rceil+1$
        &
        $0$
        &
        Lin--Shih \cite{lin1999new}, $n\geq 12$.
    \\
        $2n-2-D(n)$
        &
        $2\log(n)+O(1)$
        &
        $4$
        &
        $0$
        &
        H4 \cite{lin2003constructing}, Z4 \cite{lin2003z4}, WE4 \cite{lin2004new}, SU4 \cite{lin2005faster}.
    \\
        $2n-2-D(n)$
        &
        $\mathbf{1.4401}\log(n)+O(1)$
        &
        $D(n)+1$
        &
        $0$
        &
        Zhu--Cheng--Graham \cite{scan:zero_def_circuits06}.
    \\
        $2n-2-D(n)$
        &
        $2\log(n)+O(1)$
        &
        $\mathbf{2}$
        &
        $0$
        &
        Sheeran--Parberry \cite{sheeran2006new}, Lin--Hung \cite{lin2009straightforward}, Sergeev \cite{sergeev2024complexity}.
    \\
        $2n-2-D(n)$
        &
        $2\log(n)+O(1)$
        &
        $\mathbf{2}$
        &
        $0$
        &
        This work: Thm.\@ \ref{theorem:kronecker_circuits} for $s=2$.$^{(2)}$
    \\
        $2n-2-D(n)$
        &
        $\mathbf{1.8928}\log(n)+O(1)$
        &
        $3$
        &
        $0$
        &
        This work: Thm.\@ \ref{theorem:kronecker_circuits} for $s=3$.$^{(2)}$
    \\
    \hline
    \multicolumn{5}{l}{\footnotesize $^{(1)}$ The size and depth is reported only when $n$ is a power of two.}
    \\
    \multicolumn{5}{l}{\footnotesize $^{(2)}$ The circuit construction is LOGTIME-uniform.}
    \end{tabular}
    } 
    \label{table:previous_work}
\end{table*}
%
%
\paragraph{Contributions.}
%
In this work, we propose a new approach of constructing prefix sum circuits and algorithms by viewing the problem from a linear algebraic perspective. Specifically, recall that the prefix sum of a vector $\vecx$ can be written as $\vecy=\matL\vecx$, where $\matL$ is the lower triangular all-ones matrix. As we note in Theorem~\ref{theorem:triu_kron}, $\matL$ admits a two-term Kronecker product decomposition (see also Figure~\ref{fig:main_theorem}). This allows us to describe many different types of prefix algorithms, both existing and new ones, by exploiting the properties of Kronecker products. 
%
%
By carefully organizing the underlying operations, we obtain a new family of recursive prefix circuits, whose properties are summarized in Table \ref{table:previous_work} (see also Theorem \ref{theorem:kronecker_circuits}). These circuits are parametrized by an integer $s\in[2,n/2]$ that corresponds to the ``block-size'' of the recursion. 
We highlight the following:
%
\begin{itemize}
    \item For $s=3$, it has the lowest depth $( \approx1.893\log(n)+O(1))$ among all  zero-deficiency circuits with constant fan-out.
\end{itemize}
%
Importantly, the proposed circuits satisfy the so-called \emph{LOGTIME-uniformity}, which means that the connections between the nodes of a circuit can be verified in $O(\log(n))$ time for a sequence of length $n$ \cite{santhanam2014uniformity}. In that sense, it is the simplest zero-deficiency circuit with depth $2\log(n)+O(1)$ and fan-out two ($s=2$), and, to the best of our knowledge, it is the first LOGTIME-uniform zero-deficiency prefix circuit family with constant fan-out that achieves depth smaller than $2\log(n)+O(1)$ $(s=3)$.

As a direct application, we show how to use the proposed circuits in the context of quantum adders. Evidently, the proposed quantum adder achieves lower Toffoli depth and/or size compared to existing adders, as summarized in Table~\ref{table:quantum_adder_comparison}.
%

%
\paragraph{Notation.}
%
Matrices and vectors are denoted with capital and small Latin letters, respectively, in bold font. All vectors are considered column vectors and $\vecones_n$ is the all-ones vector with length $n$. We denote by $\matU_s$ (and $\matL_s$) the upper (and lower) triangular all-ones matrix of size $s\times s$.  $\matI_n$ is the identity matrix of size $n\times n$, and $\vece_i$ is its $i$-th column. $\matU^-_n:=\matU_n-\matI_n$, and $\matL^-_n:=\matL_n-\matI_n$ are the \emph{strictly} upper- and lower-triangular all-ones matrices, respectively. For non-negative integers $i,j,k$ we denote $[i:j]$ the set $\{i,i+1,\ldots,j\}$ and $[i:k:j]$ the set $\{i,i+k,i+2k,\ldots,i+mk\}$, where $m$ is the largest integer such that $i+mk\leq j$.
We use zero-based indexing for matrices and vectors. $\matA(i,j)$ is the element of $\matA$ in row $i$, column $j$. For a vector $\vecx$, $\vecx(i)$ is the $(i+1)$-th entry, and $\vecx(i:j)$ is the subvector $(\vecx(i),\vecx(i+1),\dots, \vecx(j))$. 
For $1<s<n$, the operator $\matt_s: \mathbb{R}^n \to \mathbb{R}^{s \times \lceil n/s\rceil}$ returns a matrix:
%
$
    \matt_s(\vecx)=\begin{pmatrix}
    \vecx_0 & \vecx_1 & \ldots & \vecx_{\lceil \frac{n}{s}\rceil}
\end{pmatrix},
$
%
where $\vecx$ is padded with zeros if $s$ does not divide $n$ and $\vecx_i:=\vecx(is:is+s-1)$. $\vect(\matX)$ stacks the columns of $\matX$ in a single column vector.
We denote $h_s(n):=\lceil\frac{n}{s}\rceil -1$, $h_s^{(k)}(n)=\underbrace{h_s( h_s(\ldots  h_s}_{k\text{ times}}(n)\ldots))$, $h_s^*(n)=\arg\max_j\left\{ h_s^{(j)}(n) > s\right\}$, and $r_s(n):=h^{(h^*_s(n))}_s(n)\in[s]$ is the ``remainder''.
%
%
\section{Kronecker products}
%
In this section, we prove that triangular all-ones matrices can be written as the sum of two Kronecker products of  triangular all-ones matrices. We first recall some useful properties of Kronecker products.

Let $\matA \in \mathbb{R}^{m \times n}$ and $\matB \in \mathbb{R}^{p \times q}$ be two matrices, and denote $a_{i,j}:=\matA(i,j)$. The Kronecker product of $\matA$ and $\matB$, denoted by $\matA \otimes \matB$, is the $mp \times nq$ block matrix:
%
\begin{align*}
\matA \otimes \matB &= 
\begin{bmatrix}
a_{0,0}\matB & a_{0,1}\matB & \cdots & a_{0,n-1}\matB \\
a_{10}\matB & a_{11}\matB & \cdots & a_{1,n-1}\matB \\
\vdots  & \vdots  & \ddots & \vdots  \\
a_{m-1,0}\matB & a_{m-1,1}\matB & \cdots & a_{m-1,n-1}\matB
\end{bmatrix}.
\end{align*}
%
We frequently use the following two Kronecker properties, see~\cite[Equations~$2.4$ and~$2.10$]{kron:mixed_prod_stats69}:
%
\begin{description}
    \item[Transpose] $(\matA \otimes \matB)^\top  = \matA^\top \otimes \matB^\top$, 
    \item[Mixed product] $
    \vect(\matA\matX\matB)=(\matB^\top\otimes\matA)\vect(\matX)$.
\end{description}
%

%
\begin{theorem}\label{theorem:triu_kron}
Fix three integers $n,n_1,n_2>1$, such that $n=n_1n_2$. The lower triangular all-ones matrix $\matL_n$ can be decomposed as follows:
\begin{align}\label{eqn:main}
    \matL_n &= \matL_{n_1}\otimes \matL_{n_2} + \matL^{-}_{n_1} \otimes                 \matU^{-}_{n_2}\nonumber\\
            &= \matI_{n_1}\otimes\matL_{n_2}+\matL^-_{n_1}\otimes \vecones_{n_2}\vecones^\top_{n_2}.
\end{align}
Equivalently, $\matU_n=\matU_{n_1}\otimes \matU_{n_2}+\matU^{-}_{n_1}\otimes \matL^{-}_{n_2}$.
\begin{proof}
To prove Equation~\eqref{eqn:main}, it suffices to show that the following equality holds:
\begin{equation}\label{eq:proof}
    \matU_n = \matU_{n_1} \otimes \vecones_{n_2}\vecones^\top_{n_2} - \matI_{n_1} \otimes \matL_{n_2}^{-}.
\end{equation}
%
Indeed,
%
\begin{align*}
    \matU_{n_1} &\otimes \vecones_{n_2}\vecones^\top_{n_2} - \matI_{n_1} \otimes \matL_{n_2}^{-} \\ & = \matU_{n_1} \otimes \left(\matU_{n_2} +\matL^{-}_{n_2} \right) - \matI_{n_1} \otimes \matL_{n_2}^{-}\\
    & = \matU_{n_1} \otimes \matU_{n_2}  + \left( \matU_{n_1} -  \matI_{n_1}\right)\otimes \matL_{n_2}^{-}\\
    & = \matU_{n_1} \otimes \matU_{n_2}  + \matU^{-}_{n_1}\otimes \matL_{n_2}^{-},
\end{align*}
%
where, in the first equality, we decomposed the all-ones matrix into its upper triangular and strictly lower triangular all-ones parts, and in the next two equalities, we rearranged terms.
Next, we prove that the right-hand side (RHS) of Equation~\eqref{eq:proof} equals $\matU_n$ by first showing that its upper triangular part is all-ones, and then, showing that its strictly lower triangular part is all-zeros.
First, notice that the matrix $\matI_{n_1} \otimes \matL_{n_2}^{-}$ is strictly lower triangular by construction. Moreover, all entries in the upper triangular part of the matrix $\matU_{n_1} \otimes \vecones_{n_2}\vecones^\top_{n_2}$ is one. These two statements imply that the entries of the upper triangular part of the RHS of Equation~\eqref{eq:proof} are equal to one.
Second, it suffices to show that the strictly lower triangular part of the matrix in the RHS of Equation~\eqref{eq:proof} is zero. By construction, all strictly lower triangular blocks of size $n_2\times n_2$ of the RHS of Equation~\eqref{eq:proof} are zero matrices, so we only need to show that all the $n_2\times n_2$ block matrices on the main diagonal of the RHS equal $\matU_{n_2}$. Consider any diagonal block of size $n_2\times n_2$ of the RHS, and notice that both Kronecker matrices share the same sizes $n_1$ and $n_2$ in their product. Hence, the first matrix summand is $\vecones_{n_2}\vecones^\top_{n_2}$ and the second matrix summand is $-\matL_{n_2}^{-}$ which gives $\vecones_{n_2}\vecones^\top_{n_2} -\matL_{n_2}^{-} = \matU_{n_2}$.
\end{proof}
\end{theorem}
%
These decompositions can also be extended to strictly upper or lower triangular all-ones matrices by subtracting the identity matrix on both sides of Equation~\eqref{eqn:main}.
%

%
Given Theorem \ref{theorem:triu_kron}, the following linear algebraic formulation of prefix sum follows. Let $\vecx$ be a vector of size $n$ with $n=n_1n_2$ as in Theorem~\ref{theorem:triu_kron}, and let $\matX=\matt_{n_2}(\vecx)
\in{\RR^{n_2\times n_1}}$. It holds that:
%
\begin{align}
    \matL_n \vecx &= \left( \matI_{n_1}\otimes\matL_{n_2}+\matL^-_{n_1}\otimes \vecones_{n_2}\vecones^\top_{n_2} \right)\vecx
    \nonumber\\
    &= \left(  \matI_{n_1}\otimes\matL_{n_2}\right)\vect(\matX) + \left( \matL^-_{n_1}\otimes \vecones_{n_2}\vecones^\top_{n_2} \right)\vect(\matX)
    \nonumber\\
    &= \vect\left(  \matL_{n_2}\matX \matI_{n_1}\right) + \vect\left( \vecones_{n_2}\vecones_{n_2}^\top \matX \matL^{-\top}_{n_1} \right)
    \nonumber
    \\
    &= \vect\left(  \matL_{n_2}\matX \right) + \vect\left( \vecones_{n_2}\vecones_{n_2}^\top \matX \matU^{-}_{n_1} \right),
    \label{eq:kronecker_scan}
\end{align}
%
where we used Theorem~\ref{theorem:triu_kron}, linearity,  and the Kronecker mixed-product property.
Interestingly, this formulation lends itself for recursive evaluations. The first term $(\matL_{n_2}\matX)$ can be evaluated as $n_1$ prefix sums of length $n_2$. The second term, $\vecones_{n_2}(\vecones_{n_2}^\top \matX)\matU^{-}_{n_1}$, is an \emph{exclusive} prefix sum of length $n_1$.   By tuning the parameters $n_1$ and $n_2$, and by choosing evaluation strategies appropriately, we can recover known algorithms and circuits (see~\Cref{table:known_algorithms_as_kronecker}), as well as to design new ones, as discussed next.
%
\begin{table}[htb]
\centering
\caption{Three examples of prefix sum algorithms that are derived by (recursively) evaluating a Kronecker product expression of $\matL_n$ or $\matU_n$.}
\label{table:known_algorithms_as_kronecker}
\setlength{\tabcolsep}{12pt} 
\renewcommand{\arraystretch}{1.1}
    \begin{tabular}{lll}
    \hline
    \textbf{Expression of $\matL_n$} & \textbf{Recursive step(s)} & \textbf{Ref.} \\
    \hline
    $\matI_{\frac{n}{2}}\otimes\matL_2+\matL^-_{\frac{n}{2}}\otimes \vecones_{2}\vecones^\top_2$ & $\left(\vecones_2^\top\matt_2(\vecx)\right)\matU_{\frac{n}{2}}$ & \cite{adders:brent_kung82} \\
    $\matI_2\otimes \matL_{\frac{n}{2}} +  \matL_2^-\otimes \vecones_{\frac{n}{2}}\vecones^\top_{\frac{n}{2}}$ & 
    $\matL_{\frac{n}{2}}\vecx_0$ and $\left(\vecones_2^\top\matt_2(\vecx_0)\right)\matU_{\frac{n}{4}}$
    & \cite{ladner1980parallel} \\
    & 
    where $\vecx_0:=\matt_{\frac{n}{2}}(\vecx)\vece_0$
    &
    \\
    $\matI_{\frac{n}{s}}\otimes\matL_s+\matL^-_{\frac{n}{s}}\otimes \vecones_{s}\vecones^\top_s$ & $\left(\vecones_s^\top\matt_s(\vecx)\right)\matU_{\frac{n}{s}}$ & \cite{scan:zouzias_europar23} \\
    $\matI_{\frac{n}{s}}\otimes\matL_s+\matL^-_{\frac{n}{s}}\otimes \vecones_{s}\vecones^\top_s$
    &
    $\left(\vecones_s^\top \matt_s(\vecx)\right)\begin{pmatrix}
    \matU_{\lceil \frac{n}{s} \rceil-1} & 0\\
    0 & 0
    \end{pmatrix}$
    &
    Thm.\@ \ref{theorem:kronecker_circuits}
    \\
    \hline
    \end{tabular}
\end{table}
%
%
%
\section{Constructing prefix circuits}
\label{section:known_algorithms_as_kronecker}
%
We start by showing that the Brent--Kung algorithm \cite{scan:brent_kung_stoc80} can be described as special cases of Eq. \eqref{eqn:main}. Here we set $n_2=2$ and $n_1=n/2$ (assuming that $n$ is even).
Brent--Kung executes the following steps. 
%
\begin{enumerate}
    \item First, it computes partial prefix sums of size $2$, i.e. $\matL_2\matX$, where $\matX=\matt_2(\vecx)\in\mathbb{R}^{2\times\frac{n}{2}}$. In linear algebra notation we write: $\begin{pmatrix}
        1 & 0\\
        1 & 1
    \end{pmatrix}\matX
    =
    \begin{pmatrix}
        \matX(0,:)\\
        \vecones_2^\top \matX
    \end{pmatrix}.
    $ 
    \item Next, the prefix sum of $\vecw=\vecones_2^\top\matX$ is computed recursively. This can be written as  $(\vecones_2^\top\matX)\matU_{\frac{n}{2}}$. 
    \item Finally, the elements of $\vecw=\vecones_2^\top\matX$ are used to complete the elements of the first 
    row of $\matX$ (except the first element). This can be written as: 
    \begin{align}
        \matX(0,1:\tfrac{n}{2}-1)+\vecw^\top(0:\tfrac{n}{2}-2).
        \label{eq:brent_kung_final_update}
    \end{align}
    \end{enumerate}
    Ultimately, the entire algorithm returns $\vect(\matY)$ where:
    \begin{align} 
    \matY=\begin{pmatrix}
        \matX(0,:)\\
        \underbrace{\underbrace{\vecones_2^\top \matX}_{\text{Step 1}} \matU_{\frac{n}{2}}}_{\text{Step 2}}
    \end{pmatrix}
    \underbrace{+
    \begin{pmatrix}
        0 & \left(\vecones_2^\top\matX\matU_{\frac{n}{2}}\right)(0:\tfrac{n}{2}-2)
        \\
        0 & \underbrace{0\quad \ldots \quad 0}_{\frac{n}{2}-1}
    \end{pmatrix}}
    _
    {\text{Step 3}}.
    \label{eq:kronecker_brent_kung}
\end{align}
%
One can verify that this is indeed a particular three-step evaluation of Eq. \eqref{eq:kronecker_scan} (left as an exercise for the interested reader). Indeed, Step 1 is the first level of the up-sweep part of Brent-Kung circuit, and Step 3 is the last level of the down-sweep part of the BK circuit. Step 2 is the recursive step.
With a bit more work, we can show that the Ladner-Fischer circuits \cite{ladner1980parallel} can also be described as a specific evaluation of Kronecker decompositions.
The number of operations in Eq. \eqref{eq:kronecker_brent_kung} is well-known. Step 1 performs $\tfrac{n}{2}$ additions in parallel. Step 2 recursively computes a prefix sum for length $\tfrac{n}{2}$. Step 3 requires $\tfrac{n}{2}-1$ operations, giving a recursive formula for the size $S(n)=n-1+S(\tfrac{n}{2})$ operations. Unrolling the recursion, we get $S(n)=2n-\log(n)-2$ when $n$ is a power-of-two. 
The corresponding recursion\footnote{Indeed, Step 1 performs additions in parallel with depth one, the second step (recursion) has depth $D(n/2)$, and the last step performs additions in parallel.} of the depth is $D(n)=1+D(n/2)+1$, which unrolls to $D(n)=2\log(n)-1$. In this scenario, the depth and sum give a suboptimal sum of $S(n)+D(n)=2n+\log(n)-3$, i.e., the resulting prefix circuit has $\log(n)+1$ deficiency.

\paragraph{Identifying the deficiency.}

\begin{figure}[htb]
    \centering
    \includegraphics[width=0.75\linewidth]{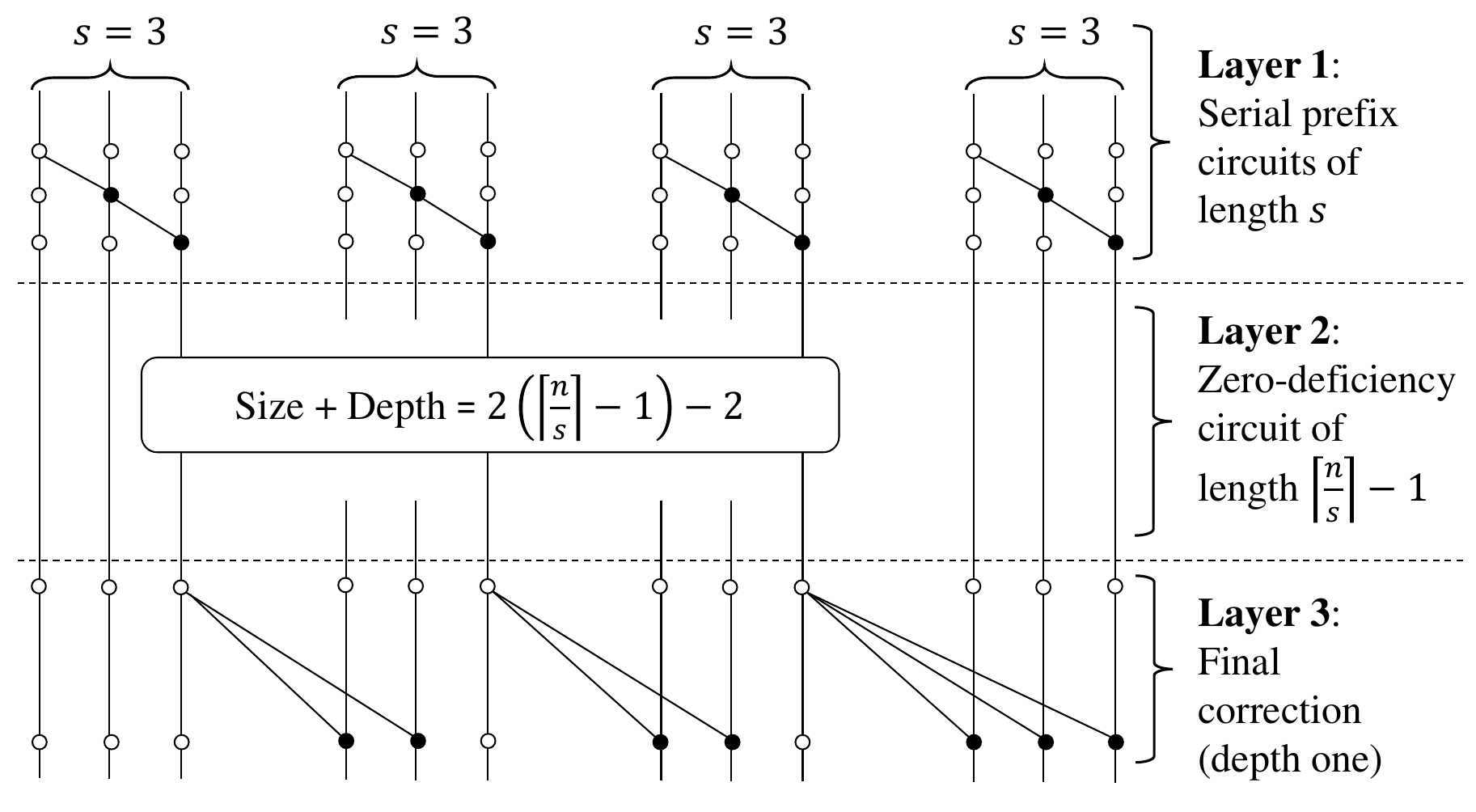}
    \caption{Zero-deficiency Kronecker prefix circuit.}
    \label{fig:simple_optimal_kronecker_circuit}
\end{figure}

Note that in the update step of Eq. \eqref{eq:brent_kung_final_update}, we only need the first $\tfrac{n}{2}-1$ elements of $1_2^\top\matX\matU_{\frac{n}{2}}$. It turns out that this is precisely the ``root'' of the deficiency: every recursive call computes one element more than what is needed! With this observation, we can modify the recursion to only compute the first $\lceil\tfrac{n}{2}\rceil-1$ elements of $1_2^\top\matX\matU_{\frac{n}{2}}$:
%
\begin{enumerate}[Step 1:]
\item $\vecw^\top\gets \vecones_2^\top\matX\in\mathbb{R}^{\lceil\frac{n}{2}\rceil}$ ($\lfloor\frac{n}{2}\rfloor$ binary additions).
\item $\vecz^\top\gets \vecw^\top(0:n')\matU_{n'}$, where $n':=\lceil\frac{n}{2}\rceil-1$.
\item Return: 
\begin{align*}\begin{pmatrix}
    \matX(0,0:n') & \matX(0,\lceil\tfrac{n}{2}\rceil)\\
    \vecz^\top(0:n') & \vecw(\lceil\tfrac{n}{2}\rceil)
    \end{pmatrix}
    +
    \begin{pmatrix}
     0 & \vecz^\top(0:n'-1) & \vecz(n')\\
     0 & 0\quad \ldots \quad 0 &
     \vecz(n')
    \end{pmatrix}.
\end{align*}
\end{enumerate}
%
Evidently, this procedure does not only remediate the Brent--Kung deficiency, but it can also be generalized for any ``block-size'' $s\geq 2$, providing a remarkably simple, zero-deficiency prefix circuit family, with low depth. 
\paragraph{A new zero-deficiency circuit family.} In Figure \ref{fig:simple_optimal_kronecker_circuit}, we illustrate the proposed family of zero-deficiency prefix circuits. It consists of three layers:
%
\begin{enumerate}
    \item The first layer partitions the $n$ inputs in blocks of size $s$, except the last block which has size $n\mod s$. A serial prefix circuit $L_i$ is used to compute the prefix of $\vecx(is:\min\{is+s,n\})$, for all $i=0,\ldots \lceil \tfrac{n}{s}\rceil-1$ simultaneously (in parallel).
    \item The second layer takes as inputs the last outputs of every $L_i$, except the last one, and computes their prefix recursively. In this layer we can actually use \emph{any} zero-deficiency circuit, and the final circuit will also have zero-deficiency. However, the depth, size, and fan-out, will be affected.
    \item In the third layer, the outputs of second layer are used to finalize the remaining partial prefixes.
\end{enumerate}
%
The analysis is summarized in Theorem \ref{theorem:kronecker_circuits}. Notably, for $s=2$, the circuit has the same $2\log(n)+O(1)$ depth as Brent--Kung and other classic circuits, but, for \emph{every} $n$, it achieves Snir's lower bound $S(n)=2n-2-D(n)$!
%
\begin{theorem}
    \label{theorem:kronecker_circuits}
    For every pair of integers $n\geq 4$ and $s\in[2,n/2]$, there exists a LOGTIME-uniform family of recursive zero-deficiency prefix circuits with depth at most $sh_s^*(n)+r_s(n)\leq s\lceil\log_s(n)\rceil-1$ and fan-out $s$.
    \begin{proof}
        We first show by induction that a recursive construction achieves zero-deficiency. In the base case, if $\lceil \tfrac{n}{s}\rceil-1\leq s$, then we simply have a single serial prefix circuit with size $n-1$ and depth $n-1$, which has zero-deficiency. For the inductive step, we argue that if the claim holds for length $\lceil \tfrac{n}{s}\rceil-1$, then it also holds for length $n$. The first layer has size $n-\lceil \tfrac{n}{s}\rceil$ and depth $s-1$. The circuit of the second layer has zero-deficiency by the induction hypothesis, and therefore its depth and size sum to $2(\lceil \tfrac{n}{s}\rceil-1)-2=2\lceil \tfrac{n}{s}\rceil -4$. The final layer has depth one and size $n-s-\lceil \tfrac{n}{s}\rceil+2$. Adding everything together, the total depth and size sum to the desired $2n-2$:
        \begin{align*}
            \overbrace{n-\lceil \tfrac{n}{s}\rceil+s-1}^{\text{Layer 1}}+\overbrace{2\lceil \tfrac{n}{s}\rceil-4}^{\text{Layer 2}}+\overbrace{n-s-\lceil \tfrac{n}{s}\rceil +3}^{\text{Layer 3}}
            =
            2n-2.
        \end{align*}
        
        We can now upper bound the depth of the recursive circuit. First, we assume that $n$ is not a power of $s$, which is addressed later. Note that after the first recursion, the number of inputs in each recursive step is never a power of $s$, even if the original $n$ is. We have that:
        \begin{align*}
            D(n) = \begin{cases}
                s+D(\lceil \tfrac{n}{s}\rceil-1), & n> s,\\
                n-1, & n\leq s.
            \end{cases}
        \end{align*}
        This quantity is always upper bounded by $s\lceil \log_s(n)\rceil-2$. To see this, note that $D(n)$ is equal to $sh^*_s(n)+r_s(n)-1$. Now, it always holds that $D(n)\leq D(s^k)$, where $k=\arg\min\{j\ |\ s^k\geq n\}$. But if we replace $n$ with $s^k$ in the above, it holds that $h^*_s(s^k)=k-1$ and $r_s(s^k)=h_s^{(k-1)}=s-1$. Therefore, the total depth satisfies $D(s^k)= s(k-1)+(s-1)-1=sk-2$. Since $k=\lceil\log_s(n)\rceil$, we finally obtain that $D(n)\leq D(s^k)=s\lceil\log_s(n)\rceil-2$. 

        So far, the maximum fan-out of any node is at most $s$. However, when $n$ is a power of $s$, there is always a single gate at the last level that has fan-out $s+1$. This can be reduced to $s$ simply by constructing a circuit for the first $n-1$ inputs (where $n-1$ is no longer a power of $s$) and then attaching a single gate that adds the $n$-th input to the $(n-1)$-th output. This increases the depth by $1$, and reduces the maximum fan-out from $s+1$ to $s$. The circuit retains zero-deficiency.

        To see that the construction can be verified in LOGTIME, note that given $n$, $s$, $i\in[n]$, $j\in [n]$, and $k\in[s\lceil \log_s(n)-1]$, as inputs, it is straightforward to answer in $O(\log(n))$ steps whether there is a connection between node $i$ at level $k$ and node $j$ at level $k+1$.
\end{proof}
\end{theorem}
%
%
%
%
\paragraph{Optimizing the block size.}
%
For $s=2$, we obtain a zero-deficiency circuit with depth at most $2\lceil\log(n)\rceil-1\in 2\log(n)+O(1)$, which is the same as Brent--Kung for powers of two. However, $s=2$ does not always give the minimum achievable depth. Indeed, for larger values of $n$ we can sharpen the bound by setting $s=3$, which gives $D(n)\leq 3\lceil\log_3(n)\rceil-2\in \frac{3}{\log(3)}\log(n)+O(1)\approx 1.8928\log(n)+O(1)$. For any $n$, the actual minimum depth can be computed with dynamic programming: $\min D(n)=\min_{2\leq s \leq n/2}\{s+\min D(\lceil\frac{n}{s}\rceil-1)\}$.
%
\section{Quantum adders}
\begin{figure}[htb]
    \centering
    \includegraphics[width=0.75\linewidth]{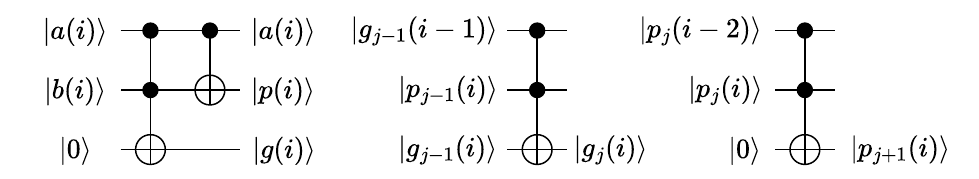}
    \caption{Toffoli gates for carry propagation/generation.}
    \label{fig:toffoli_gates}
\end{figure}

%
Here we mention how to apply the proposed prefix circuits to quantum adders. Such circuits have received significant attention over the last three decades since they serve as building blocks for important problems such as discrete logarithm and integer factoring \cite{shor1994algorithms,beckman1996efficient,takahashi2010quantum,ruiz2017quantum,thapliyal2021quantum,takahashi2008fast,draper2006logarithmic,wang2025optimal}. 
Here we follow the methodology of \cite{draper2006logarithmic,wang2025optimal}, where the main idea is to use a prefix circuit for carry propagation/generation (carry look-ahead framework). Circuits are evaluated on the following metrics:
\begin{description}
    \item[Toffoli count] The total number of Toffoli gates.
    \item[Toffoli depth] The maximum number of Toffoli gates in a path from an input to an output.
    \item[Qubit count] The total number of auxiliary qubits.
\end{description}
The goal is to minimize the Toffoli depth/count by carefully overlapping different independent Toffoli layers, while still using as few auxiliary qubits as possible. Figure \ref{fig:toffoli_gates} shows three cases where Toffoli gates are used in the quantum circuit for propagation and generation. We refer also to \cite[Sections 3 and 4]{wang2025optimal} for a broad overview of the techniques and details of the specific operations.
%
%
%
\begin{figure}[htb]\renewcommand{\baselinestretch}{0.9}
    \scriptsize
    \hrulefill\\
    \raggedright \textbf{KroneckerQuantumAdder}$(n,s)$:
    \begin{enumerate}[$\quad$]
        \item $1: \vecg(i)\gets \mathcal{T}\big(\veca(i), \vecb(i),\ket{0}\big),$\hfill $i\in[0:1:n-1]$.
        \item $*: \vecp_{0}(i):=\vecb(i)\gets
        \veca(i)\oplus \vecb(i),$ \hfill $i\in[0:1:n-1]$.
        \item $2: \vecp_1(i)\gets \mathcal{T}\big(\vecp_0(i-1),\vecp_0(i),\ket{0}\big),$\hfill$i\in[s+1:s:n-1].$
        \item $*: \vecz\gets\ket{0}_{n-s}$. \hfill\texttt{/* Aux.\@ qubit register */}
        \item \textbf{Recurse}$\big(n,s,\vecp_0,\vecp_1,\vecg\big).$
        \item $3:$ Uncompute $\vecp_1$.
        \item $4:$ Uncompute $\vecp_0$ and finalize sum.
    \end{enumerate}
    \ \\
    \raggedright \textbf{Recurse}$\big(n,s,\vecp_{t-1},\vecp_{t},\vecg\big)$:
    \begin{enumerate}[$\quad$]
        \item\textbf{if} $n\leq s:$ \hfill\texttt{ /* Final Layer */}
        \begin{enumerate}[$\quad$]
            \item\textbf{for} $i=1,\ldots,n-1:$
            \begin{enumerate}[$\quad$]
                \item $i: \vecg(i)\gets \mathcal{T}\Big(
                    \vecg(i-1),
                \vecp_{t-1}(i),
                \vecg(i)
                \Big)$
            \end{enumerate}
        \end{enumerate}
        \item \textbf{else}:
        \begin{enumerate}[$\quad$]
            \item Set $n'=\left\lceil\tfrac{n}{s}\right\rceil -1$ and $I=\left[s-1:s:n'-1\right]$.
            \item $*: \vecp_{t+1}\gets \ket{0}_{n'}$. \hfill\texttt{/* Aux.\@ qubit register */}
            \item  \textbf{Serial}$\big(n,s,\vecp_{t-1},\vecp_{t},\vecp_{t+1},\vecg\big)$.
            \item  \textbf{Recurse}$\big(n', s,\vecp_{t}(I), \vecp_{t+1}(I), \vecg(I)\big)$.
            \item  \textbf{Finalize}$\big(n, s, \vecp_{t}, \vecp_{t+1}, \vecg\big)$.
        \end{enumerate}
    \end{enumerate}
    \ \\
    \raggedright \textbf{Serial}$\big(n,s,\vecp_{t-1},\vecp_{t},\vecp_{t+1},\vecg\big)$:
    \begin{enumerate}[$\quad$]
        \item\textbf{for} $k=1,\ldots,s-2:$
        \begin{enumerate}[$\quad$]
            \item \textbf{for} $i\in[k+s:s:n-1]:$
            \begin{enumerate}[$\quad$]
                \item $k: \vecg(i)\gets
                \mathcal{T}
                \Big(
                \vecg(i-1),
                \vecp_{t-1}(i),
                \vecg(i)
                \Big)$
                \item $k: \vecp_{t}(i+1)
                \gets
                \mathcal{T}
                \Big(
                \vecp_{t-1}(i), \vecp_{t}(i+1),
                \ket{0}
                \Big)$
            \end{enumerate}
        \end{enumerate}
        \item Set $k=s-1$
        \item $k: \vecg(i)\gets 
        \mathcal{T}
        \Big(\vecg(i-1),
        \vecp_{t-1}(i),
        \vecg(i)
        \Big)$\hfill $i\in[s-1:s:n-1]$
        \item $k: \vecp_{t+1}(i)\gets 
        \mathcal{T}\Big(
        \vecp_{t}(i-s),
        \vecp_{t}(i),
        \ket{0}
        \Big)$ \hfill $i\in[(s-1)^2:s^2:n-1]$
    \end{enumerate}
    \ \\
    \raggedright \textbf{Finalize}$\left(n, s, \vecp_{t}, \vecp_{t+1}, \vecg\right)$:
    \begin{enumerate}[$\quad$]
        \item\textbf{for} $i\in [s:s:n-2]:$
        \begin{enumerate}[$\quad$]  
        \item\textbf{for} $k\in \big[i:\min\{i+s-2,n-2\}\big]:$
            \begin{enumerate}[$\quad$]
                \item $*:$ $\vecz(is+k)\gets\vecg(i-1)\oplus \vecz(is+k)$ 
                \item $1: \vecg(k)\gets
                \mathcal{T}\Big(
                \vecz(is+k), 
                \vecp_{t}(k), 
                \vecg(k)
                \Big)$.  
            \end{enumerate}
        \end{enumerate}
    \item 
    $1: \vecg(n-1)\gets
    \mathcal{T}
    \Big(
    \vecg(s\lfloor\tfrac{n}{s}\rfloor), 
    \vecp_{t}(n-1),
    \vecg(n-1)
    \Big)$.
    \item $1: $ Uncompute $\vecp_{t+1}$ and $\vecz$.
    \end{enumerate}
    \hrulefill
    \caption{Algorithm to construct quantum adder. The lines that are numbered indicate the corresponding Toffoli layer (layers marked as $*:$ do not contain Toffoli gates).}
    \label{figure:quantum_adder_new_algorithm}
\end{figure}
%
%
\begin{table}[htb]
\small
    \centering
    \caption{Comparison of known quantum adders with respect to Toffoli depth, count, and number of auxiliary qubits. BK:=Brent--Kung, Other $\in$ \{Sklansky, Kogge--Stone, Han--Carlson, Ladner--Fischer\}.}
    \setlength{\tabcolsep}{10pt} 
    \begin{tabular}{l r r r}\hline
    Quantum Adder &  Toffoli Count & Toffoli Depth & Qubit Count \\\hline\hline
        BK \cite{draper2006logarithmic,wang2025optimal}     & $O(n)$ & $2\log(n)+O(1)$ & $O(n)$  \\
        Other, Strategy 1 \cite{wang2025optimal}  & $O(n\log(n))$ & $2\log(n)+O(1)$ & $O(n\log(n))$  \\
        Other, Strategy 2 \cite{wang2025optimal}  & $O(n\log(n))$ & $\log(n)+O(1)$  & $O(n\log(n))$  \\
        Theorem \ref{theorem:quantum_adder},  $s=3$ & $O(n)$ & $1.893\log(n)+O(1)$ & $O(n)$ \\
        \hline
    \end{tabular}
    \label{table:quantum_adder_comparison}
\end{table}
%
During the final correction step in each recursion, the maximum fan-out can be as large as $s+1$. A standard way to circumvent this is to use a layer of CNOT gates to replicate the corresponding output to $s+1$ $\ket{0}$-registers (see e.g. \cite{wang2025optimal}). Since the CNOT gates do not contribute to the Toffoli size/depth of the circuit, we omit this step in the algorithm description for simplicity. 
In Theorem \ref{theorem:quantum_adder} we summarize the analysis of the proposed quantum adder, and in Table \ref{table:quantum_adder_comparison} we compare the evaluation metrics with existing circuits (the results are imported from \cite{wang2025optimal}). Algorithm \ref{figure:quantum_adder_new_algorithm} describes the procedure to construct the corresponding quantum circuit, where $\mathcal{T}(\ket{a},\ket{b},\ket{c}):=\left(a\cdot b\right)\oplus c$ is a Toffoli gate operating on qubits $a,b,$ and $c$. Below we provide the proof of the Theorem.
%
\begin{theorem}
    \label{theorem:quantum_adder}
    Let $\ket{\veca}$ and $\ket{\vecb}$ be two $n$-qubit integers. We can prepare a quantum circuit which computes the sum $\ket{\veca}+\ket{\vecb}$ that has at most $s\lceil \log_s(n)\rceil+2$ depth, $O(n)$ Toffoli gates, and $O(n)$ auxiliary qubits. For $s=3$ the Toffoli depth is $\approx 1.893\log(n)+O(1)$.
    \begin{proof}
        We start bounding the qubit count. The algorithm uses $n-s$ auxiliary qubits for the register $\vecz$ and at most $O(n/s^t)$ qubits for each $\vecp_t$, for $t=1,2,\ldots,\lceil\log_s(n)\rceil+1$. This gives a total of $O(n)$ auxiliary qubits.
        
        For the Toffoli depth, there are two layers of Toffoli gates before the first recursive call, and two layers after (for uncomputation). $\textbf{Recurse}(\cdot)$ has Toffoli depth at most $s\lceil\log_s(n)\rceil-2$, which is the depth of the Kronecker prefix circuit. Therefore, the total Toffoli depth is at most $4+s\lceil\log_s(n)\rceil-2=s\lceil\log_s(n)\rceil+2$.

        The total number of Toffoli gates is bounded as follows. In the first two propagation layers, $\vecp_{0}$ and $\vecp_{1}$, there are a total of $3n/2$ gates, and additional $n/2$ gates for the un-computation of $\vecp_{1}$. In each recursive step $t=1,\ldots,\lceil\log_s(n)\rceil-1$, there are at most $n/s^t$ gates for $\vecg^{(t+1)}$, at most $n/s^{t+1}$ gates for $\vecp^{(t+2)}$, at most $n/2^t$ gates for $\vecg^{(t)}$, and finally at most $n/2^{t+1}$ gates for the uncomputation of the $\vecp^{(t+2)}$. This gives a total of at most $3n/2^{t}$ gates at level $t$. Summing for all $t=1,\ldots,\lceil\log(n)\rceil-1$ gives at most $3n$ Toffoli gates.
\end{proof}
\end{theorem}
%
%
\section*{Acknowledgements}
%
We thank Igor Sergeev for helpful discussions and for indicating important references.
%
\section{Conclusion}
%
In this work we revisited algorithms and circuits for the prefix sum problem, through the lens of linear algebra. By decomposing triangular all-ones matrices as the sum of two Kronecker products, we were able to describe a new family of recursive zero-deficiency prefix circuits. These circuits are parametrized by an integer $s$ (the block-size). By choosing $s$ appropriately, we can obtain circuits with reduced depth and/or fan-out compared to existing zero-deficiency families. As an application, these techniques were used to construct quantum adders with reduced Toffoli depth and/or size, compared to existing ones.
%

%
\end{document}